\documentclass[epjST]{svjour}

\usepackage{graphicx}
\usepackage{bm}
\usepackage{hyperref}
\usepackage{amsmath}
\usepackage{braket}
\usepackage{dsfont}	
\usepackage{siunitx}	

\newcommand{\ee}{\mathrm{e}}
\newcommand{\ii}{\mathrm{i}}

\usepackage{cite}
\begin{document}

\title{High-order harmonic generation in hexagonal nanoribbons}

\author{Christoph J\"ur{\ss}\thanks{\email{christoph.juerss@uni-rostock.de}} \and Dieter Bauer\thanks{\email{dieter.bauer@uni-rostock.de}}}

\institute{Institute of Physics, University of Rostock, 18051 Rostock, Germany}

\abstract{
 The generation of high-order harmonics in finite, hexagonal nanoribbons is simulated. Ribbons with armchair and zig-zag edges are investigated by using a tight-binding approach with only nearest neighbor hopping. By turning an alternating on-site potential off or on, the system describes for example graphene or hexagonal boron nitride, respectively. The incoming laser pulse is linearly polarized along the ribbons. The emitted light has a polarization component parallel to the polarization of the incoming  field. The presence or absence of a polarization component perpendicular to the polarization of the incoming field can be explained by the symmetry of the ribbons. Characteristic features in the harmonic spectra for the finite ribbons are analyzed with the help of the band structure for the corresponding periodic systems. 
} 

\maketitle

\section{Introduction} 
	
	Ultrafast dynamics in condensed matter systems have been studied intensively in recent years \cite{Ghimire2011, SchubertO.2014,  Ndabashimiye2016, LangerF.2017, TancPhysRevLett.118.087403,Zhang2018,Vampa2018,Garg2018}. In particular, high-order harmonic generation (HHG) has proven to be a powerful tool   as it is able to probe static and dynamic properties of the solid target by all-optical means \cite{VampaPhysRevLett.115.193603,Hohenleutner2015,Luu2015,You2017,Baudisch2018}.
	
	HHG was initially observed for atoms and molecules in the gas phase.  For non-perturbative laser intensities and photon energies well below the ionization potential, the energy of the emitted photons can be large multiples of the incident photon's energy, the high-order harmonics. 
	
	The mechanisms underlying HHG in solids are similar to those in the gas phase. For instance, the  celebrated  semi-classical three-step model \cite{corkum_plasma_1993,LewensteinPhysRevA.49.2117} introduced for isolated atoms, where, in the first step,  the electron is excited into the continuum. In the second step, the electron propagates in the presence of the laser-field, and, in the third step, it recombines with the  ion  upon generating   a photon with an energy given by the kinetic energy of the electron at the time of recombination and the ionization potential.  A similar model exists for solids \cite{Vampa_theory_2014,vampa_merge_2017} where, first,  the electron is excited from the valence band to the conduction band, second, the electron in the conduction band and the hole in the valence band propagate in the presence of the laser field, and, third, the electron and hole recombine upon generating a harmonic photon.
	
	If the solid is an insulator or semi-conductor, it has a non-vanishing band gap between the valence and the conduction band. 
	The three-step model of solid state HHG provides a way to separate two different contributions of harmonic radiation \cite{Vampa_theory_2014}. First, the movement of the the electron (and hole) inside the bands create intraband harmonics. As the electron and and hole recombine, a transition between both bands occur. The radiation from this transition is called interband harmonics.

    Many studies focus on the bulk of a solid. In reality, solids are finite and have edges. Edge states might cause interesting effects in high-harmonic spectra, in particular when they are topological in nature  \cite{bauer_high-harmonic_2018,DrueekeBauer2019,JuerssBauer2019}. In this paper, we  focus on the high-harmonic spectra from finite systems and compare with the corresponding result for the bulk. We restrict ourselves to the topologically trivial phase in this work. 
    
    Graphene is one particularly interesting two-dimensional solid because of its relativistic Dirac cones. In graphene, the atoms form a hexagonal lattice structure. Hexagonal boron nitride (h-BN) is a different example with the same lattice structure. HHG in hexagonal lattice structures has been studied for the bulk and for ribbons for topologically trivial graphene and h-BN \cite{PhysRevB.95.035405,Chizhova_2017,Yoshikawa2017,Hafez2018,Yue2020} and the topologically nontrivial Haldane model \cite{Silva2019,chacon_observing_2018,juerss2020helicity}.  

	In this work, we investigate the generation of high-harmonics in hexagonal nanoribbons for two different edge configurations: zig-zag and armchair. Ribbons with and without alternating on-site potentials are investigated in the topologically trivial phase only. The system without alternating on-site potential contains one atomic element (as, e.g., in graphene) whereas two different elements are contained in the case of alternating on-site potential (e.g., h-BN).  
	
	The outline of this work is as follows. In Chapter \ref{sec:theory}, the basic theory is summarized, starting with the finite system without external field in Sec.\ \ref{sec:static}. We show the calculation and the results for the band structures for the armchair (Sec.\ \ref{sec:static_arm}) and  the zig-zag (Sec.\ \ref{sec:static_zz}) ribbon with periodic boundary conditions. In Sec.\ \ref{sec:coupling_field} the coupling to an external field is presented. HHG spectra for the finite armchair (Sec. \ref{sec:res_arm}) and finite zig-zag (Sec. \ref{sec:res_zz}) ribbons are discussed.
	
	\section{Theory}\label{sec:theory}
		
\begin{figure} 
			\includegraphics[width = \columnwidth]{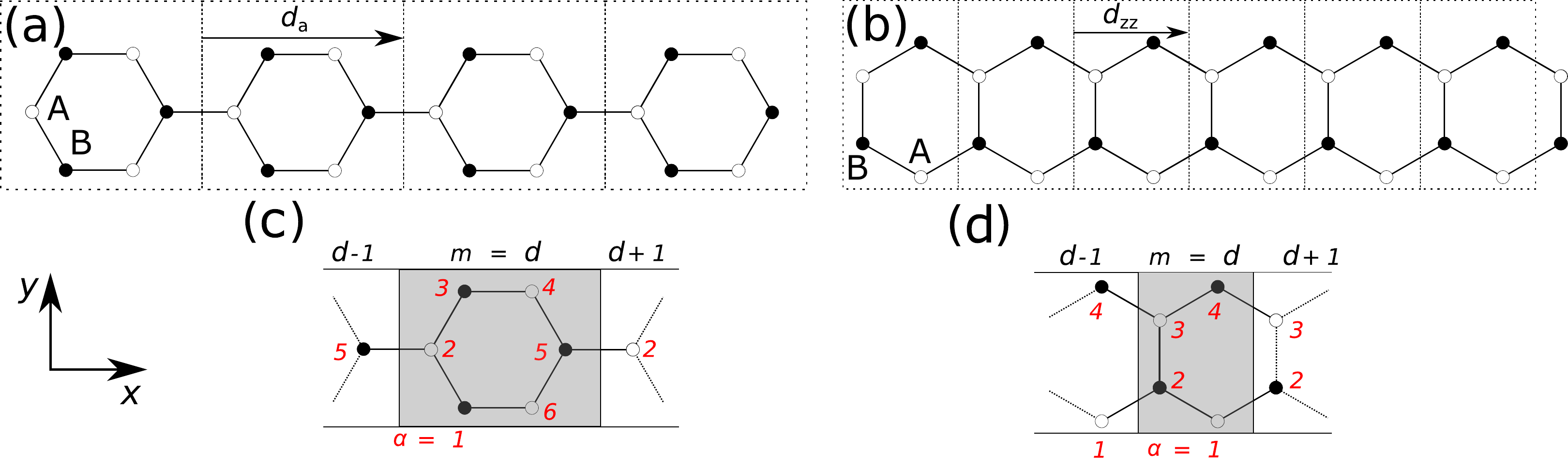}
			\caption{\label{fig1} Sketch of ribbons built from hexagons, for finite (a,b) and  infinite (c,d) ribbons. (a,c) armchair and (b,d) zig-zag configuration. The distance between nearest neighbors (indicated by solid lines) is $a$, the hopping amplitude between them is $t_1
			\in\mathds{R}$. An alternating on-site potential $M$ ($-M$) at sites A (B), indicated by unfilled (filled) circles, is included. The unit cells are marked by dotted rectangles. The lattice constant is $d_\mathrm{a}$ for the armchair and $d_\mathrm{zz}$ for the zig-zag ribbon. Here, the finite armchair ribbon contains $N_\mathrm{hex} = 4$, the zig-zag $N_\mathrm{hex} = 6$ hexagons (unit cells). For the infinite ribbon (c,d), the hoppings inside a unit cell $m = n$ and to neighboring unit cells $m = n\pm 1$ are indicated by solid lines. The lattice sites inside a unit cell are labeled by $\alpha$ (red).} 
		\end{figure}

		In this work, we investigate hexagonal ribbons  in two different configurations, armchair (Fig. \ref{fig1} (a)) and zig-zag (Fig. \ref{fig1} (b)).  We consider two different types of sites: A and B. The on-site potential is $M$ ($-M$) on lattice sites A (B). In Fig. \ref{fig1}, the lattice sites A and B are indicated by unfilled and filled circles, respectively. The lattice constant for the armchair ribbon is given by $d_\mathrm{a} = 3 \,a$, where $a$ is the distance between two neighboring sites. For the zig-zag ribbon, the lattice constant is $d_\mathrm{zz} = \sqrt{3}\,a$. Atomic units (a.u.), $\hbar=|e|=m_e=4\pi\epsilon_0=1$, are used if not stated otherwise.
		
		\subsection{Static system}\label{sec:static}
		    
		    The systems have $N$ atomic sites. The atomic orbital at site $i$ is denoted as $\ket{i}$. A general single-electron wavefunction  is given by
		    \begin{equation}
		        \ket{\psi} = \sum_{i = 1}^N g_i\ket{i}.
		    \end{equation}
		    The Hamiltonian in position space and tight-binding approximation reads
		  \begin{equation}\label{eq:t_indep_H}
		        \hat{H}_0 = t_1\sum_{<i,j>}
		        \left( \ket{j}\bra{i} + \mathrm{h.c.}\right) + M\left(\sum_{i\in A} \ket{i}\bra{i} - \sum_{i\in B} \ket{i}\bra{i}\right)
		    \end{equation}
		    where the sum $\sum_{<i,j>}$ runs over nearest neighbors $i$ and $j$ and the sums $\sum_{i\in A,B}$ over sites $A$ or $B$, respectively. The parameter $t_1$ is the hopping amplitude between adjacent sites. Hopping between next-nearest neighbors is not considered in this work.  The eigenstates $\ket{\psi_i}$ fulfill the time-independent Schr\"odinger equation (TISE)
		    \begin{equation}
		        \hat{H}_0\ket{\psi_i} = E_i\ket{\psi_i}.
		    \end{equation}
		
		    In the following, we describe the propagation of states $\ket{\psi}$ in position space  for ribbons with $N_\mathrm{hex}$ hexagons (Fig. \ref{fig1} (a,b)) in time. However, one aim of this work is to relate features in the harmonic spectrum of the finite ribbons with energy differences in  the band structure of the corresponding ribbon bulk. Hence, band structures are calculated for the ribbons with periodic boundary conditions in $x$-direction, see Fig. \ref{fig1} (c,d). The resulting Hamiltonian will be given in crystal-momentum space ($k$-space).
		
		    For the distance between adjacent sites we take the value for  graphene \cite{Cooper_2012}, i.e., $a=2.68$\,a.u.$\simeq 1.42$\,\AA.
		    For the nearest-neighbor hopping amplitude we use data from simulations without tight-binding approximation \cite{Drueeke_SI} with which we want to compare. To that end the energies of an armchair ribbon with four unit cells were calculated, and the nearest-neighbor hopping was adjusted till the band gap was identical for both methods. As a result we find $t_1 = -0.07776 \approx -2.116~\mathrm{eV}$. The negative sign of $t_1$ is chosen to obey the node rule of quantum mechanics, as in this case the coefficients $g_i$ have all the same sign for the state with the lowest energy.

		    \subsubsection{Armchair ribbon}\label{sec:static_arm}
		    
\begin{figure} 
        			\includegraphics[width = \columnwidth]{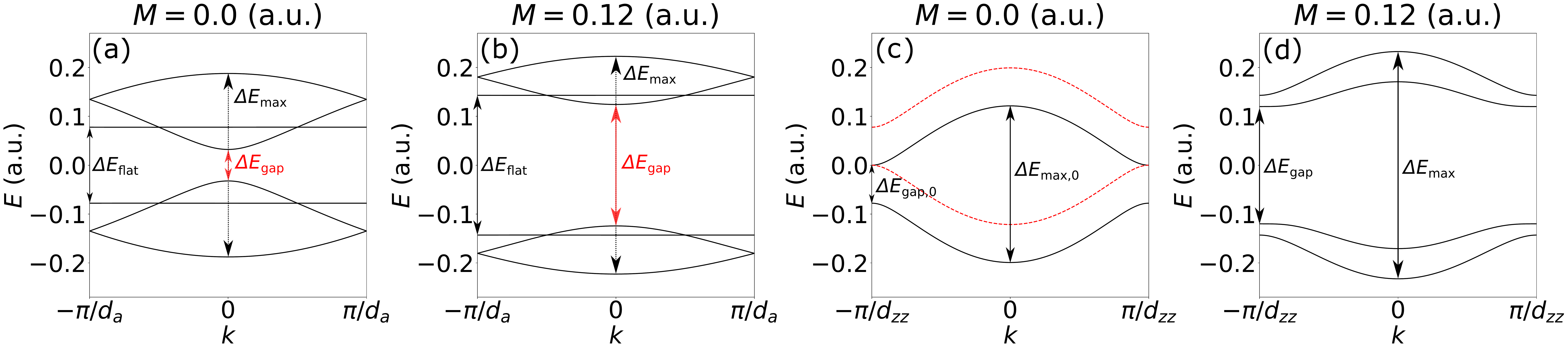}
        			\caption{\label{fig2} Band structure for the bulk of the ribbon with armchair (a,b) and zig-zag (c,d) edges for $M = 0$ (a,c) and $M = 0.12$ (b,d) in the first Brillouin zone.}  
        		\end{figure}

                The calculation leading to the bulk-Hamiltonian for the armchair ribbon is  in Appendix \ref{app:arm_bulk}. The unit cell contains six sites, see Fig. \ref{fig1} (c). Hence, there are six bands, see Fig. \ref{fig2} (a,b). The energies are calculated numerically.
                
                The armchair ribbon has a band gap of $\Delta E_\mathrm{gap} = 0.08338$ for a vanishing on-site potential $M = 0$ (Fig. \ref{fig2} (a)). 
                The band gap $\Delta E_\mathrm{gap}$ increases with the on-site potential $M$ (Fig. \ref{fig2} (b)). The band structure is symmetric around $E = 0$. Two of the bands are flat, their energy is constant over the whole Brillouin zone. It is given by $E_\mathrm{flat} = \pm\sqrt{t_1^2 + M^2}$, see appendix \ref{app:arm_bulk}.
                
                In the finite ribbon with $N_\mathrm{hex}$ hexagons, there are also $N_\mathrm{hex}$ states that have the same energy. Their energy is identical to the energy of the flat bands in the bulk.
                
		     \subsubsection{Zig-zag ribbon}\label{sec:static_zz}
		        The bulk-Hamiltonian for  the zig-zag ribbon was calculated in \cite{juerss2020helicity}. 
		        Here, we do not consider hopping between next-nearest neighbors as in \cite{juerss2020helicity} (i.e., $t_2 = 0$). The bulk-Hamiltonian reads
		        \begin{align}\label{eq:Ham_bulk_zz}
		            \hat{H}_\mathrm{bulk,zz}(k) = \begin{pmatrix}
		            M& T_1(k) & 0&0\\
		            T_1(k_i)&-M&t_1&0\\
		            0 & t_1 & M&T_1(k)\\
		            0&0&T_1(k)&-M
		            \end{pmatrix}
		        \end{align}
		        with $T_1 = 2 t_1 \cos (k_i d_\mathrm{zz}/2)$. The TISE for the bulk is given by 
		        \begin{align}
		            \hat{H}_{\mathrm{bulk,zz}}\bm{u}(k) = E(k)\bm{u}(k)
		        \end{align}
		        with the periodic factor  $\bm{u}(k) = \left(u_1(k),u_2(k),u_3(k),u_4(k)\right)^\top$ in the Bloch-like  ansatz. 
		       
        		Other than for the armchair ribbon, the energies of the zig-zag ribbon can be written in a compact, analytical form
        		\begin{align}
        		    E(k) = \pm\sqrt{M^2 + t_1^2/4\left(\sqrt{16\,\mathrm{cos}^2\left(k\,d_\mathrm{zz}/2\right) + 1 } \pm 1\right)^2},
        		\end{align}
        		where both $\pm$ are independent, leading to four bands. Results with and without $M$ are shown in Fig. \ref{fig2} (c,d). For a vanishing on-site potential (c) there is no band gap between the bands with a negative energy (valance bands) and the bands with a positive energy (conduction bands). However, only transitions between the two black, solid or the two red, dashed bands are allowed for a linearly polarized laser field in dipole approximation  \cite{Hsu_2007,Saroka_2017}. This creates an effective band gap, $\Delta E_\mathrm{gap,0}$. 
        		
        		A band gap centered at $E = 0$ appears for non-vanishing  on-site potential. It is given by $\Delta E_\mathrm{gap} = 2|M|$, an example is shown in  Fig. \ref{fig2} (d). Transitions between all bands are allowed for $M\neq 0$.

        \subsection{Coupling to an external field}\label{sec:coupling_field}
		
		    The coupling of the systems to an external field and the  propagation of an electronic wavefunction in time  is described in Ref.\  \cite{juerss2020helicity}. 
		    
		    The vector potential is linearly polarized along the $x$-direction (i.e., along the ribbons). For times $0\leq t \leq 2\pi n_{cyc}/\omega_0$, the vector potential is given by
		    \begin{align}
    		    \bm{A}(t) = A_0~\sin ^2\left(\frac{\omega_0 t}{2 n_{cyc}}\right)~\sin (\omega_0 t) \bm{e}_x,
    		\end{align}
    		and it is zero otherwise.  The following laser parameters are used if not stated otherwise: amplitude $A_0 = 0.2$ (intensity  $\simeq 7.9\times 10^{10}~ \mathrm{Wcm}^{-2}$), angular frequency $\omega_0 = 7.5\cdot 10^{-3}$ (i.e., wavelength $\lambda = \SI{6.1}{\micro\meter}$), and the pulse comprises  $n_{cyc} = 4$ cycles.  
		    
		    The total current is given by
		    \begin{align}
			    \bm{J}(t) = \sum_{l}\bra{\Psi^l(t)}\hat{\bm{j}}(t)\ket{\Psi^l(t)},
		    \end{align}
		    i.e., the sum over all currents arising from the occupied states $\ket{\Psi^l(t)}$, propagated in time.
		    The current operator reads \cite{Review_Transport}
    		\begin{align}\label{eq:current_operator}
    			\hat{\bm{j}}(t) = 	-\mathrm{i}\sum_{i,j}\left(\bm{r}_{i} - \bm{r}_{j} \right)\ket{i}\bra{i}H(t)\ket{j}\bra{j},
    		\end{align}
		    with $\bm{r}_{i,j}$ being the position of the sites $i, j$ with their respective orbitals $\ket{i}$ and $\ket{j}$ and $H(t)$ the time-dependent Hamiltonian (see Ref. \cite{juerss2020helicity}).
		    It is assumed that at the beginning of the pulse, all eigenstates with negative energy (i.e., below the Fermi level) are occupied.
		    
		    The intensity of the emitted light is proportional to 
		    \begin{align}
		        \big|P_{\parallel,\perp}(\omega)\big|^2 = \left|\mathrm{FFT}\left[\dot{J}_{x,y}(t)\right]\right|^2.
		    \end{align}
		    The symbols $\parallel$ and $\perp$ denote the parallel ($x$-direction) and perpendicular ($y$-direction) polarization direction with respect to the polarization direction of the linear polarized laser pulse.

	\section{Results}\label{sec:results}

	    In this paper, we discuss the high-order harmonic spectra for an armchair ribbon consisting of $N_\mathrm{hex} = 4$ hexagons ($N = 24$) and a zig-zag ribbon built of $N_\mathrm{hex} = 6$ hexagons ($N = 26$). The results are compared with simulations without the tight-binding approximation for systems of the same size \cite{Drueeke_SI}. We briefly discuss the size-dependence of the zig-zag ribbon at the end of this Section.

	    \subsection{Ribbon with armchair edges}\label{sec:res_arm}
	    
	    \begin{figure} 
			\includegraphics[width = \columnwidth]{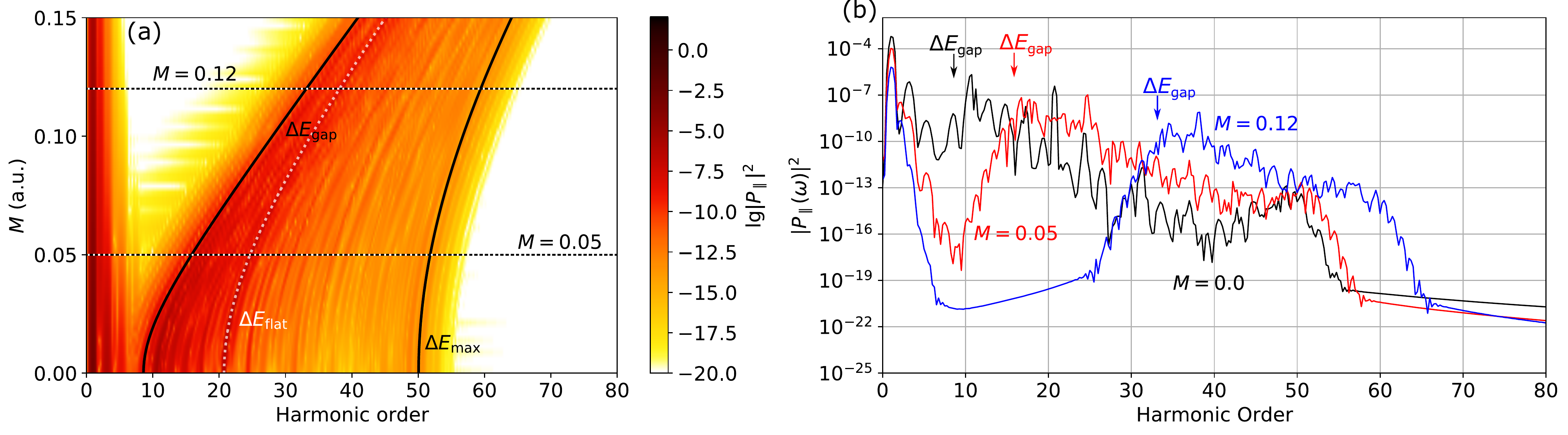}
			\caption{\label{fig3} (a) High-order harmonic spectra for a finite armchair ribbon with $N_{\mathrm{hex}} = 4$ hexagons as function of the on-site potential $M$. The spectra  show the emission polarized parallel to the  polarization direction of the incoming laser field. The line representing $\Delta E_\mathrm{gap}$ ($\Delta E_\mathrm{max}$) indicates the minimal band gap (maximal energy difference) between the valence and conduction bands of the respective periodic system. The horizontal lines mark the on-site potential of the spectra shown in (b).}
		\end{figure}

	    The high-order harmonic spectra for parallel polarization direction as function of the on-site potential for the armchair ribbon are shown in Fig. \ref{fig3} (a). In addition, the spectra for $M = 0$,  $M =0.05$, and $M = 0.12$ are shown in Fig. \ref{fig3} (b). The energy is given in units of the laser frequency $\omega_0 = 0.0075$ (i.e.,  harmonic order). The laser field is polarized linearly along the ribbon. Light with a polarization direction perpendicular to the field is not emitted. This is due to the symmetry of the system in that direction (i.e., the $y$-direction) even with a non-vanishing on-site potential (see Fig. \ref{fig1} (a)). In both plots, the minimal band gaps of the periodic system between valence and conduction band $\Delta E_\mathrm{gap}$ are indicated. The band gap increases with the on-site potential $M$. The line $\Delta E_\mathrm{max}$ (Fig.\  \ref{fig3} (a)) shows the maximal energy difference between valence and conduction band. It also increases with $M$. The horizontal lines in Fig.\  \ref{fig3} (a) mark those $M$s for which spectra are shown in Fig. \ref{fig3} (b). 
	    
	    The band gap of the periodic system with vanishing on-site potential is $E_\mathrm{gap} = 0.0644 \approx 8.6~\omega_0$.
	    For the finite system with $N_\mathrm{hex} = 4$ one finds a band gap of $0.08338 \approx 11.1~ \omega_0$. The band gap of the finite system is larger because there are only $24$ eigenstates of the Hamiltonian. Due to the restricted number of states, the sampling of the energy spectrum is not sufficient to capture the minimal band gap of the periodic ribbon. A larger finite chain would resolve it  but is not part of this study. We refer to Ref. \cite{PhysRevA.97.043424}, where the size dependency of a one-dimensional, linear chain was studied.
	    For an on-site potential of $M = 0.12$, the band gap of the finite system is given by $0.245 \approx 33.9~\omega_0 $ and that  of the periodic system is $E_\mathrm{gap} = 0.248 = 33.1~\omega_0 $. Here, the band gap of the finite and the band gap of the periodic system are similar.
	    
	    In the harmonic spectra, one can see that the harmonic yield for small energies drops exponentially. Up to the energy of the band gap, the harmonic yield is relatively low. For energies larger than the band gap, one observes a plateau where the yield  is almost constant. An ultimate  cut-off is observed at an energy that corresponds to the maximum energy difference between valence and conduction bands $\Delta E_\mathrm{max}$. Transitions with larger energies are not possible in the tight-binding model. Hence, no harmonics are emitted at larger energies. 
	    
	    The harmonics below the band gap are dominated by movement of electrons inside the bands \cite{Navarrete_kRes_2019}, known as intraband harmonics. Due to the fully occupied valence bands, there are electrons that move in opposite directions because of opposite band curvature, and therefore the emitted radiation by "individual" electrons destructively interferes, leading to the drop in the harmonic yield \cite{bauer_high-harmonic_2018}. The plateau for larger energies is dominated by transitions between the valence and conduction bands \cite{Navarrete_kRes_2019}, called interband harmonics. The band gap increases with the on-site potential $M$. As a consequence, the smallest energy of the plateau-region shifts to higher harmonic orders. The ultimate cut-off of the plateau also increases, because the maximal energy difference of the bands also becomes larger with increasing $M$. The onset and the cut-off of the plateau can be estimated by the periodic system. Its minimal band gap $E_\mathrm{gap}$ and maximal energy difference $E_\mathrm{max}$ is plotted in Fig. \ref{fig3} (a).  The colored-contour plot in Fig.\ \ref{fig3} (a) might not be able to show the starting and beginning of the plateau properly. It is better visible in Fig. \ref{fig3} (b). The colored vertical arrows indicate the band gap of the respective periodic system. It agrees with the onset of the plateau if the on-site potential $M$ is non-zero. The cut-off of the plateau for $M = 0$ is at around harmonic order $50$ and for $M = 0.12$ at around $60$ harmonic orders. The maximal energy difference of the periodic system is given by $\Delta E_\mathrm{max} = 0.375\approx 50.1~\omega_0$ and $\Delta E_\mathrm{max} = 0.446 \approx 59.4~\omega_0$ for $M = 0$ and $M = 0.12$, respectively, agreeing  with the cut-offs.
	    The plateau is restricted to the region between  $E_\mathrm{gap}$ and $E_\mathrm{max}$, as expected.
	   
	    The flat bands of the band structure (see Fig. \ref{fig2} (a,b)) are separated by an energy of $\Delta E_\mathrm{flat} = 2 |t_1| = 0.1556 \approx 20.7~\omega_0$ for $M = 0$. The harmonic spectrum shows a peak at this energy. With increasing on-site potential $M$, the energy difference between those bands increases, indicated by the dotted line $\Delta E_\mathrm{flat}$ in  Fig. \ref{fig3} (a). There are as many states with the same energy inside the flat bands as there are hexagons in the ribbon. Therefore, many possible transitions have the same transition energy. This large number of transitions with identical energies causes the peak in the spectrum. 
		
		\begin{figure} 
			\includegraphics[width = \columnwidth]{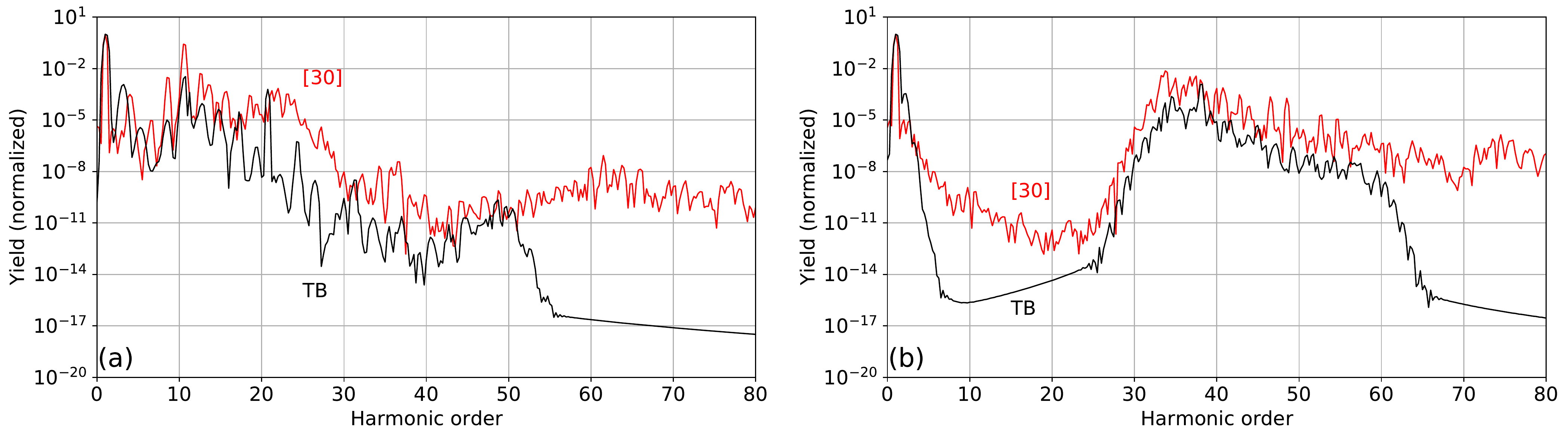}
			\caption{\label{fig3_2} Comparison of the harmonic yield (normalized to the maximal harmonic yield) of the armchair ribbon for the tight-binding model used in this work (TB) and the calculation without tight-binding approximation from Ref. \cite{Drueeke_SI}. For (a) the on-site potential is $M = 0.0$ ($V_\mathrm{os} = 0.0$) and for (b) it is $M = 0.12$ ($V_\mathrm{os} = 0.4$). $V_\mathrm{os}$ the used parameter of the on-site potential in the reference.}
		\end{figure}

	    The results presented so far are qualitatively the same as the results in Ref.\  \cite{Drueeke_SI}, in which no tight-binding approximation is used. This can be seen from Fig. \ref{fig3_2}. The system with an on-site potential of $M = 0.12$ has approximately the same band gap as the system with an on-site potential of $V_\mathrm{os} = 0.4$ in that reference  (Fig. \ref{fig3_2} (b)). The method without the tight-binding approximation includes states above the conduction bands. 
	    Therefore, the spectra in that reference also show harmonics with larger energies than $\Delta E_\mathrm{max}$. These harmonics are absent when using the tight-binding approximation, here one should include more bands in order to obtain the correct spectra. However, the suppressed harmonic yield below the band gap is visible and also the slope of the plateau is similar. The advantage of the tight-binding approximation is the computational time. The algorithm here is approximately three orders of magnitude faster than the one in Ref. \cite{Drueeke_SI}.
	    
	    Note that the hopping parameter $t_1$ is chosen in order to fit the band gap of the systems for both methods. However, the maximal energy difference $E_\mathrm{max}$ is different. One reason for that is the symmetry of the tight-binding bulk Hamiltonian that enforces mirror-symmetric valence and conduction bands about the zero-energy axis. This symmetry is absent in the continuous description of Ref.\ \cite{Drueeke_SI}.

	    \subsection{Ribbon with zig-zag edges}\label{sec:res_zz}

\begin{figure} 
			\includegraphics[width = \columnwidth]{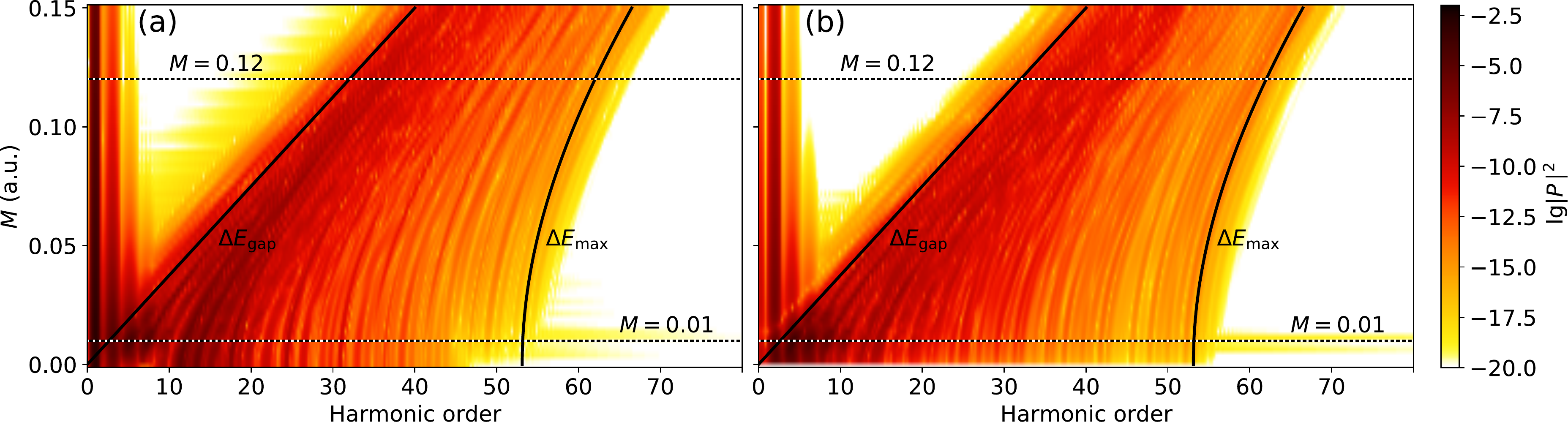}
			\caption{\label{fig4} Harmonic spectra for the finite zig-zag ribbon containing $N_\mathrm{hex} = 6$ hexagons as function of the on-site potential $M$. Spectra in parallel (a) and perpendicular (b) polarization direction to the incoming field are shown. The line $\Delta E_\mathrm{gap}$ indicates the minimal band gap, the line $\Delta E_\mathrm{max}$ the maximal gap between valence and conduction band as function of $M$ for the periodic system.}  
		\end{figure}

\begin{figure}
		    \centering
			\includegraphics[width = \columnwidth]{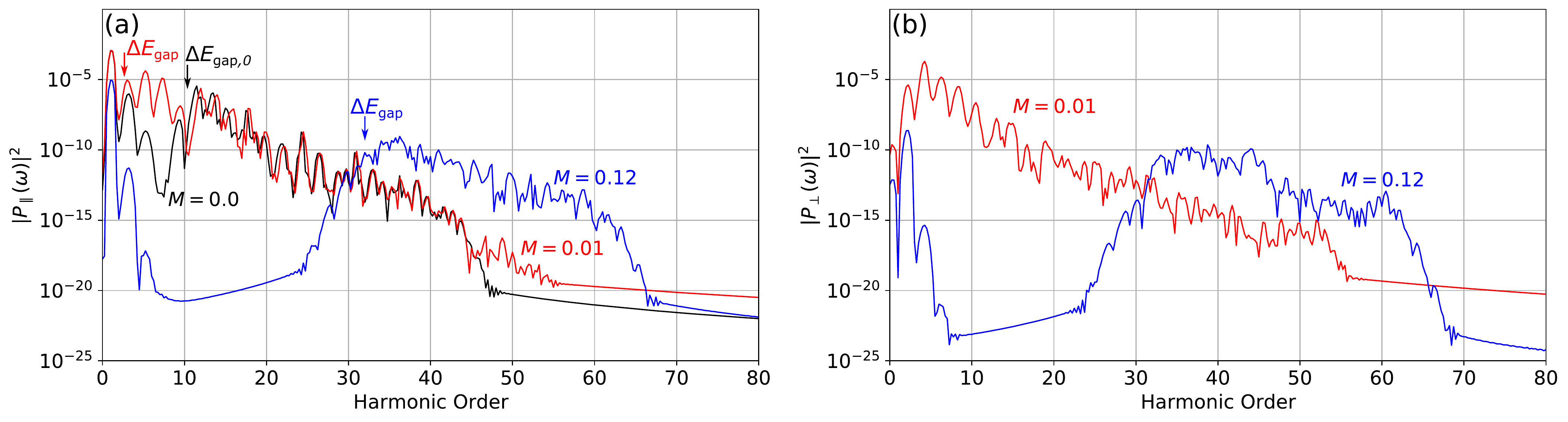}
			\caption{\label{fig5} High-order harmonic spectra of a finite zig-zag ribbon with $N_{\mathrm{hex}} = 6$ hexagons for different on-site potentials $M$  for the parallel (a) and perpendicular (b) polarization direction to the incoming field. The vertical arrows mark the energy of the band gap for the respective periodic system.}  
		\end{figure}

		Harmonic spectra for the finite zig-zag ribbon with $N_\mathrm{hex} = 6$ hexagons as function of $M$ are shown  in Fig. \ref{fig4} in parallel (\ref{fig4} (a)) and perpendicular (\ref{fig4} (b)) polarization direction with respect  to the polarization of the incoming laser field. In addition, Fig. \ref{fig5} shows spectra for three different on-site potentials $M$ (Fig. \ref{fig5} (a) in  parallel and Fig. \ref{fig5} (b) in perpendicular polarization direction). The corresponding $M$ values  are indicated by horizontal lines in Fig. \ref{fig4}. The marked energies $\Delta E_\mathrm{gap}$ and $\Delta E_\mathrm{max}$ indicate the minimal band gap and the maximal energy difference between valence and conduction band of the periodic system, respectively. 
		
		In both figures, one can see that without on-site potential, the zig-zag ribbon does not emit light perpendicular to the polarization direction of the incoming laser field (i.e., the $y$-direction). As the on-site potential becomes finite, the symmetry of the system in $y$-direction is broken. This can be seen in Fig. \ref{fig1} (b): the on-site potential at the lowest sites ($\alpha = 1$) is  $M$, on the topmost sites  ($\alpha = 4$) it is $-M$. As a consequence, the electrons are attracted more towards the upper sites than to the lower sites. Hence, light polarized in  $y$-direction is now also emitted, i.e.,  perpendicular to the polarization of the  incoming laser field, see Figs. \ref{fig4} (b) and \ref{fig5} (b).
		
		The band gap increases linearly with the on-site potential  $M$  and vanishes for $M = 0$. The spectra without on-site potential show an exponential decrease of the harmonic yield. A typical drop similar to the one for the armchair ribbon can be observed below harmonic order 11, best seen in Fig. \ref{fig5} (a). This drop of the harmonic yield is explainable by the destructive interference of the intraband emission. One could expect that  the interband harmonics should compensate the drop in the yield because of the vanishing band gap. However, as was shown in Refs.\  \cite{Hsu_2007,Saroka_2017}, transitions between certain bands are forbidden in the graphene zig-zag ribbon. This fact was already indicated in Fig. \ref{fig2} (c): transitions are only allowed between the lowest valence and the lowest conduction band (black, solid lines) and between the highest valence and conduction band (red, dashed lines). Therefore, the effective minimal band gap of the periodic system is given by $\Delta E_{\mathrm{gap},0} = |t_1| = 0.07776 \approx 10.4 ~\omega_0$. This is in good agreement with the onset of the plateau. The maximal energy difference between the bands where transitions are allowed is given by $\Delta E_{\mathrm{max},0} = 0.321 \approx 42.7~ \omega_0$, which agrees well with the  cut-off.

	    Further, as the band gap increases, we can see the typical drop of the harmonic yield for energies below $\Delta E_{\mathrm{gap}}$. The plateau lies in an energy region between $\Delta E_{\mathrm{gap}}$ and $\Delta E_{\mathrm{max}}$ for both polarization directions. This shows that the overall qualitative features in the harmonic spectra of this small zig-zag nanoribbon   can be already understood  with the help of the band structure of the  periodic system.
		The results are similar to simulations without tight-binding approximation \cite{Drueeke_SI}. The only difference is the presence of harmonics above $\Delta E_{\mathrm{max}}$ without tight-binding approximation due to higher lying states, similar to the armchair ribbon.
	    
	    As the on-site potential increases, the selection rule for $M=0$ does not apply anymore due to the broken symmetry in $y$-direction. 
		However, we observe for the small system with   $N_\mathrm{hex} = 6$ and a small on-site potential that the spectra still show a drop in the harmonic yield for small energies, indicating the destructive interference of the intraband harmonics and an onset of the interband plateau  only at higher harmonic order than expected from $\Delta E_{\mathrm{gap}}$. Hence,  transitions between the highest valence band and the  lowest conduction band are still  very unlikely in such a small, finite zig-zag ribbon, otherwise the interband harmonics would fill up  the drop of the intraband harmonic yield. 
		
\begin{figure} 
		\centering
			\includegraphics[width = 0.5\columnwidth]{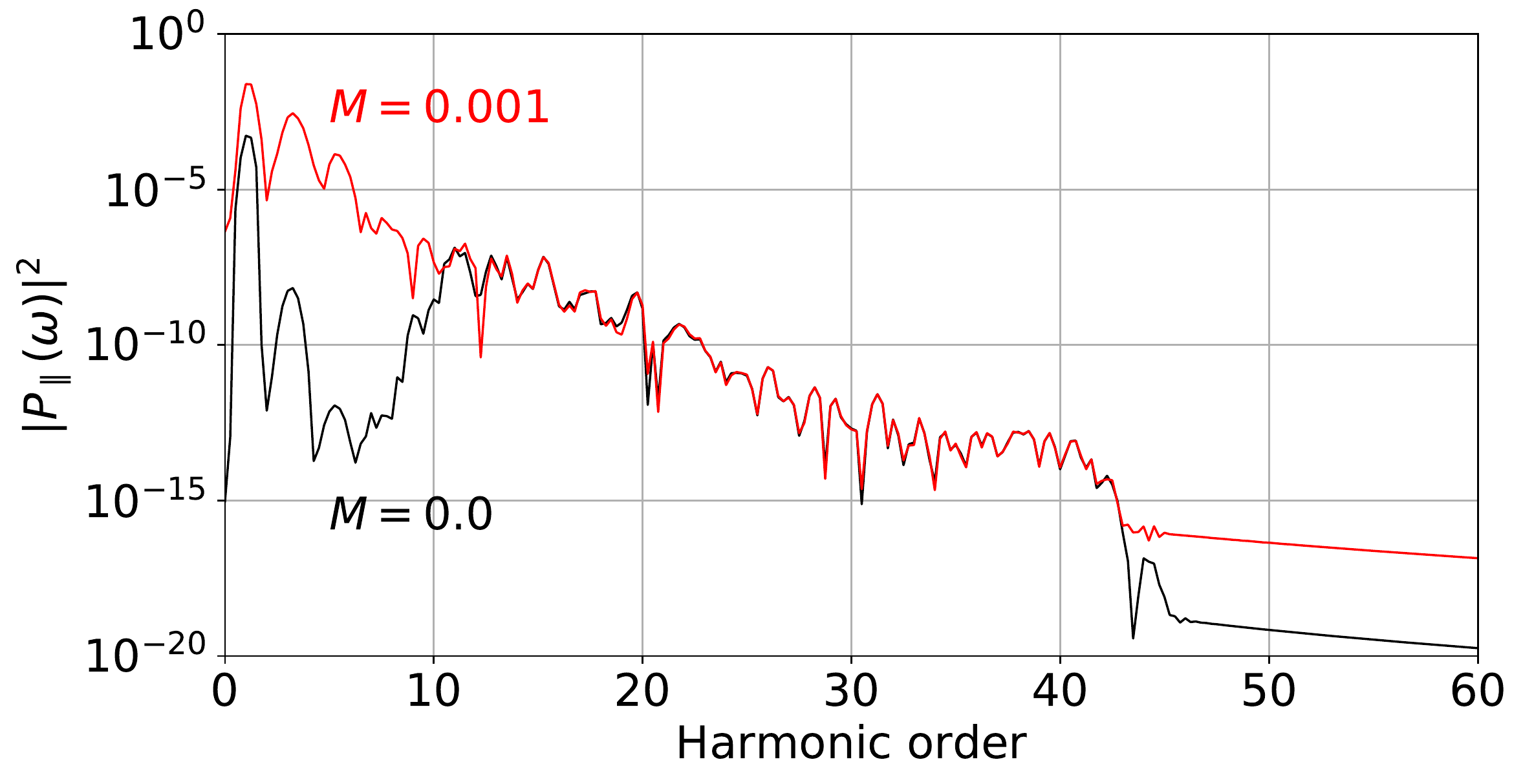}
			\caption{\label{fig6} High-harmonic spectrum in parallel polarization direction to the incoming field for a zig-zag ribbon with $N_\mathrm{hex} = 15$ hexagons for on-site potential $M=0$ and $M=0.001$.  The amplitude of the vector potential is $A_0 = 0.05$.}  
		\end{figure}

		The approximation of a finite system by a periodic system fits better for larger finite systems. In Fig. \ref{fig6}, the harmonic spectrum in parallel polarization direction of a finite chain containing $N_\mathrm{hex} = 15$ hexagons is shown. For a vanishing on-site potential $M = 0.0$, the drop in the harmonic yield up to order $11$ can be observed.

		For a small on-site potential of $M = 0.001$ the harmonic yield below the band gap is increased by several orders. Still, both valence bands are fully occupied, which means that the intraband harmonics interfere destructively as for $M = 0.0$ due to the movement of the electrons inside the bands. Obviously the interband harmonics compensate the dropping intraband harmonic yield and thus the minimal energy of the interband harmonics must be close to zero. This is only possible if transitions between all bands are allowed, showing that for longer, finite zig-zag ribbons the $M=0$ selection rule breaks more abruptly for slightly non-vanishing $M$ than for shorter ribbons.   
		
		We note that in the  calculation for $N_\mathrm{hex} = 15$ we chose the amplitude of the vector potential $A_0 = 0.05$ (i.e., intensity $\simeq 7.9\times 10^{10}~ \mathrm{Wcm}^{-2} $). For the same intensity as before ($A_0 = 0.2$) the yield drop for $M = 0$ is not  clearly visible. With increasing laser intensity the excursion of the electrons along the crystal momentum increases, diminishing the   destructive interference of the intraband emission.  As a result, the drop in the harmonic yield is not as pronounced as for smaller intensities. The fact that in the small ribbon with $N_\mathrm{hex} = 6$   the drop is more pronounced  might be due to the smaller number of states that do not resemble well continuous bands. 
		
	\section{Summary and outlook} \label{sec:summandout}
	
        In this work, we simulated high-harmonic generation in finite hexagonal nanoribbons with armchair and zig-zag edges. In an intense laser field polarized linearly along the ribbon, the armchair ribbon emits linearly polarized light parallel to the polarization of the incoming field.
        The zig-zag ribbon emits light parallel and perpendicular to the polarization of the  incoming  laser pulse  if an alternating on-site potential is included.
        Both ribbons show a suppressed harmonic yield for energies below the band gap. The band gap itself is determined by the on-site potential. The main result of this manuscript is that characteristic features in the harmonic spectra, such as onset and cut-off of the interband-harmonics plateau,  can be understood with the help of the band structure for the corresponding periodic systems.  The results for the finite ribbons are similar to those from simulations without the tight-binding approximation.
	
	\section*{Acknowledgment}
	
	C.J. acknowledges financial support by the doctoral fellowship program of the University of Rostock.

    \section*{Author contribution statement}
    C.J. performed the numerical simulations, analyzed the results, and wrote the manuscript. D.B. provided critical feedback, supported the analysis of the results, and improved the final version of the manuscript.
    
    \begin{appendix}
    
    \section{Derivation of armchair bulk-Hamiltonian}\label{app:arm_bulk}
    
        For an armchair ribbon with $N$ unit cells and periodic boundary conditions, the Hamiltonian reads
        \begin{align}
            \hat{H}_0^\mathrm{arm} &= \sum_{m = 1}^N\left[\sum_{\alpha = 1}^6 \left(t_1\ket{m,\alpha}\bra{m,(\alpha + 1) \,\mathrm{mod} \, 6} + \left(-1\right)^{\alpha + 1} \frac{M}{2}\ket{m,\alpha}\bra{m,\alpha}\right)\right.\nonumber\\
            &\left.+ t_1\ket{m,5}\bra{m+1,2}   \right] + \mathrm{h.c.},
        \end{align}
        where we now write the state $\ket{i}$ at site $i$ as $\ket{m,\alpha}$ where $\alpha$ indicates the  site within unit cell $m$. In order to obtain the bulk-Hamiltonian, we make a Bloch-like ansatz, taking the relative position within a unit cell into account, 
        \begin{align}
            \ket{\psi(k)} &= \frac{1}{\sqrt{N}}\sum_{m = 1}^N \ee^{\ii m k d_a}\ket{m}\otimes \left(u_1(k)\ee^{\ii k d_a/6}\ket{1} + u_2(k)\ket{2}+ u_3(k)\ee^{\ii k d_a/6}\ket{3} \right.\nonumber\\
            &  \left.+u_4(k)\ee^{\ii k d_a/2}\ket{4} + u_5(k)\ee^{2\ii k d_a/3}\ket{5} + u_6(k)\ee^{\ii k d_a/2}\ket{6}\right).
        \end{align}
        Here,  $d_a = 3 a$ is the lattice constant for the armchair ribbon.
        We plug this ansatz into the time-independent Schr{\"o}dinger equation, $\hat{H}_0\ket{\psi(k)} = E(k)\ket{\psi(k)}$, and multiply by $\sqrt{N}\ee^{-\ii m' k d_a}\bra{m'}$ from the left, leading to
        \begin{align}
            \hat{H}_{\mathrm{bulk, arm}}\bm{u}(k) = E(k)\bm{u}(k),
        \end{align}
        with
        \begin{align}
            \hat{H}_\mathrm{bulk,arm} = \begin{pmatrix}
		M & t_1 \ee^{-\ii k d_a/6} & 0 & 0 & 0 & t_1 \ee^{\ii k d_a/3}\\
		t_1 \ee^{\ii k d_a/6} & -M & t_1 \ee^{\ii k d_a/6} & 0 & t_1 \ee^{-\ii k d_a/3} & 0\\
		0 & t_1 \ee^{-\ii k d_a/6} & M & t_1 \ee^{\ii k d_a/3} & 0 & 0\\
		0 & 0 & t_1 \ee^{-\ii k d_a/3} & -M & t_1 \ee^{\ii k d_a/6} & 0\\
		0 & t_1 \ee^{\ii k d_a/3} & 0 & t_1 \ee^{-\ii k d_a/6} & M & t_1 \ee^{-\ii k d_a/6}\\
		t_1 \ee^{-\ii k d_a/3} & 0 & 0 & 0 & t_1 \ee^{\ii k d_a/6} & -M 
		\end{pmatrix},
        \end{align}
        where $\bm{u}(k) = \left( u_1(k),u_2(k),u_3(k),u_4(k),u_5(k),u_6(k)\right)$.
        For given $k$, one obtains six eigenstates $\bm{u}^j(k)$ and energies $E^j(k)$ ($j = 1,2,\ldots,6$). Hence, the system has six bands in the tight-binding approximation. We solve the eigenvalue equations numerically for each $k$. However, we show the calculation of one band analytically. For the periodic part of the Bloch-state we make the ansatz 
        \begin{align}\label{eq:ansatz_flat}
            \bm{u}_\mathrm{flat}(k) = \left(\beta,0,-\beta,-\gamma,0,\gamma\right)^\top,
        \end{align}
        where the values of $\beta$ and $\gamma$ are unknown and $k$-dependent. Note that this state is zero at the connection points $\alpha=2$ and $5$  to the neighboring  hexagons (see Fig.\ \ref{fig1}(c)). We obtain
        \begin{align}
            \hat{H}_\mathrm{bulk,arm}\,\bm{u}_\mathrm{flat}(k) = \begin{pmatrix}
	       M\,\beta +t_1\,\ee^{\ii k d_a/3}\,\gamma\\0\\-M\,\beta - t_1\,\ee^{\ii k d_a/3}\,\gamma\\M\,\gamma  - t_1\,\ee^{-\ii k d_a/3}\,\beta\\0\\-M\,\gamma + t_1\,\ee^{-\ii k d_a/3}\,\beta \end{pmatrix}\overset{!}{=} E_\mathrm{flat}\left(\beta,0,-\beta,-\gamma,0,\gamma\right)^\top.
        \end{align}
        This relation holds if 
        \begin{align}
            M\,\beta +t_1\,\ee^{\ii k d_a/3}\,\gamma = E_\mathrm{flat} \beta ~~~~\mathrm{and}~~~~
            M\,\gamma - t_1\,\ee^{-\ii k d_a/3}\,\beta = E_\mathrm{flat} \gamma,
        \end{align}
        resulting in the energy
        \begin{align}\label{eq:energy_flat}
            E_\mathrm{flat} = \pm\sqrt{t_1^2 + M^2}.
        \end{align}
        As this energy is independent of $k$, the corresponding two  bands  are flat. 
        
        One can also show that the same energies are obtained for a finite armchair ribbon containing $N_\mathrm{hex}$ hexagons. The ansatz is that the state is zero everywhere except at one hexagon, where it is given by eq. (\ref{eq:ansatz_flat}). The same energy (\ref{eq:energy_flat}) is obtained. The degeneracy is given by the number of hexagons in the ribbon  $N_\mathrm{hex}$. 
        
    \end{appendix}

	\bibliography{biblio}

\end{document}